\documentclass[prl,twocolumn]{revtex4}
\usepackage{graphicx}
\usepackage{amssymb,amsmath,bm}
\usepackage{enumerate}

\newcommand{\be}{\begin{eqnarray}}
\newcommand{\ee}{\end{eqnarray}}

\begin{document}

\title{Distributed chaos and inertial ranges in turbulence}

\author{A. Bershadskii}

\affiliation{
ICAR, P.O. Box 31155, Jerusalem 91000, Israel
}

\begin{abstract}

It is shown that appearance of inertial range of scales, adjacent to distributed chaos range, results in adiabatic invariance of an energy correlation integral for isotropic homogeneous turbulence and for buoyancy driven turbulence (with stable or unstable stratification, including Rayleigh-Taylor mixing zone). Power spectrum of velocity field for distributed chaos dominated by this adiabatic invariant has a stretched exponential form $\propto \exp(-k/k_{\beta})^{3/5}$. Results of recent direct numerical simulations have been used in order to support these conclusions.

\end{abstract}

\maketitle

\section{Introduction}
Exponential spectrum is a typical characteristic of smooth deterministic chaos \cite{fm},\cite{sig}. Isotropic and homogeneous turbulence emerges from distributed chaos \cite{b1} with spectrum 
$$
E(k ) \simeq \int_0^{\infty} \mathcal{P} (\kappa)~ e^{-(k/\kappa)}  d\kappa \eqno{(1)}
$$
where  $\mathcal{P}(\kappa )$ is a distribution of the wavenumber $\kappa$ of the waves driving the chaos. An asymptotic theory, developed in the Ref. \cite{b1}, relates the dispersion asymptotic scaling of the driving waves velocity
$$
\upsilon (\kappa )\propto \kappa^{\alpha} \eqno{(2)}
$$ 
to the stretched exponential spectrum of the distributed chaos
$$
E (k) \propto \exp(-k/k_{\beta})^{\beta}  \eqno{(3)}
$$
where
$$
\beta =\frac{2\alpha}{1+2\alpha}   \eqno{(4)}
$$ 
\begin{figure}
\begin{center}
\includegraphics[width=8cm \vspace{-1cm}]{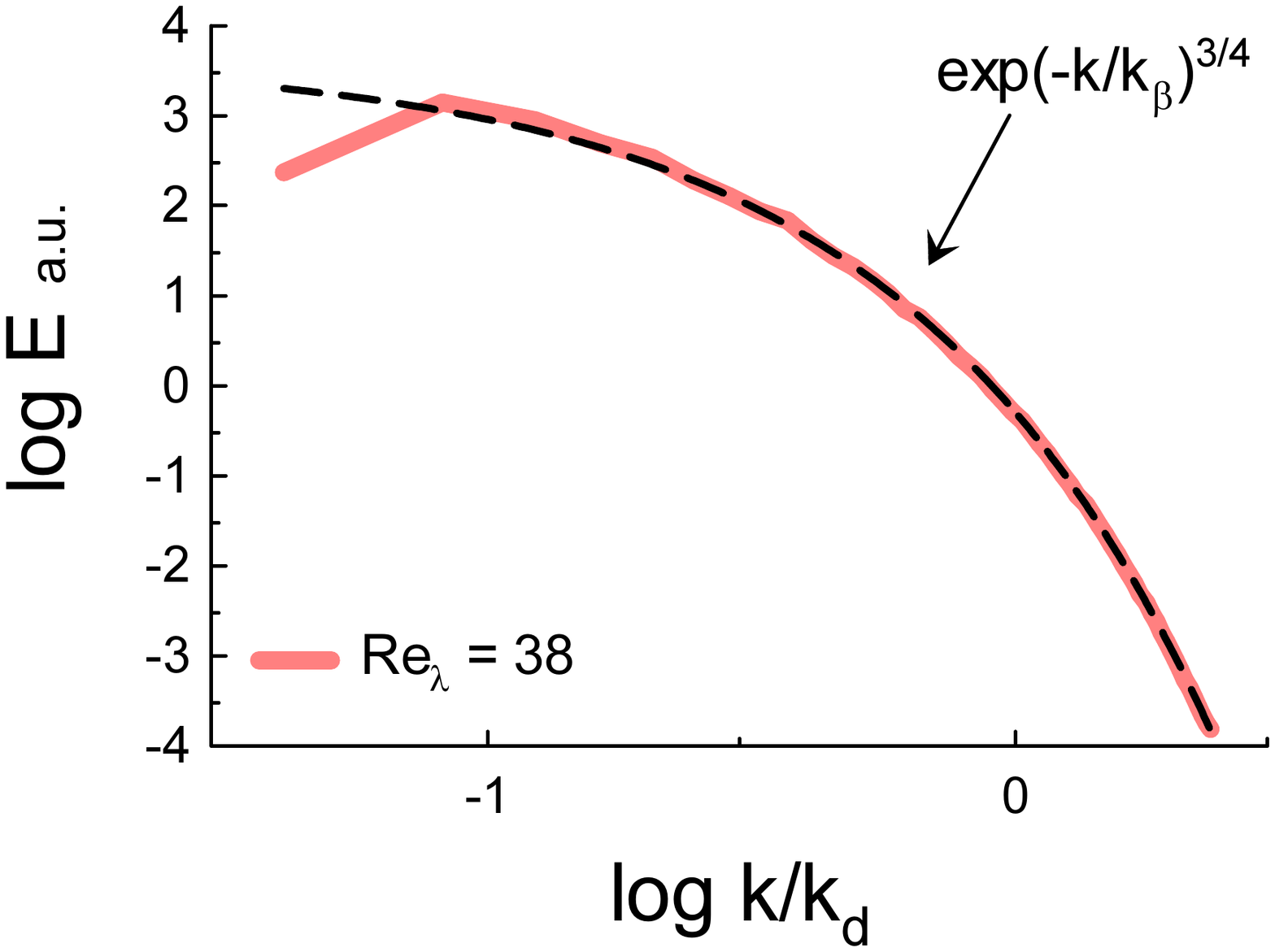}\vspace{-3.5cm}
\caption{\label{fig1} 3D energy spectrum obtained in a numerical simulation of the isotropic homogeneous turbulence \cite{gfn} at Reynolds number $Re_{\lambda} =38$. The dashed line is drawn in order to indicate the distributed chaos spectrum Eq. (3) with the $\beta = 3/4$.}
\end{center}
\end{figure}
For the isotropic homogeneous turbulence (without inertial range of scales) the space translational symmetry - homogeneity, dominates the scaling Eq. (2). Due to the Noether's 
theorem \cite{ll2} this symmetry results in the momentum conservation, represented by the momentum correlation integral
$$
I_2 =\int  \langle {\bf u} \cdot  {\bf u'} \rangle d{\bf r} \eqno{(5)}
$$   
invariance \cite{saf}-\cite{dav2}. Here ${\bf u} = {\bf u} ({\bf x},t)$ is the velocity field and ${\bf u'} ={\bf u} ({\bf x} + {\bf r},t) $. This integral is also called Birkhoff-Saffman invariant. Substituting $I_2$ into Eq. (2) and using the dimensional considerations one obtains
$$
\upsilon (\kappa )\propto I_2^{1/2}~\kappa^{3/2} \eqno{(6)}
$$
Then using Eq. (4) one obtains $\beta =3/4$. 

    Figure 1 shows 3D energy spectrum obtained in a numerical simulation of the isotropic homogeneous turbulence \cite{gfn} at Reynolds number $Re_{\lambda} =38$ (in the log-log scales, $k_d$ is the Kolmogorov's scale \cite{my}). The dashed line is drawn in order to indicate the distributed chaos spectrum Eq. (3) with the $\beta = 3/4$. For more examples and details see the Ref. \cite{b1}.  \\

\section{Inertial range domination}

   This is valid for the smooth systems. For sufficiently large Reynolds numbers the isotropic homogeneous turbulence becomes not smooth (rough) at certain region of scales. Usually the wavenumbers corresponding to such 'rough' range are smaller then those corresponding to the distributed chaos range. The rough range can produce scaling spectra and it is related to a new attractor or even to a new solution of the nonlinear equations \cite{b}.  A prominent example of such range is known as an inertial one, because for its scales the inertial (nonlinear) terms becomes dominating. Let us, following to a recent Ref. \cite{br}, estimate relative contribution of the different component (advection and pressure) of the nonlinear part of the 
Navier-Stokes equations   
$$   
\frac{\partial \bm u(\bm x,t)}{\partial t} + (\bm u(\bm x,t)\cdot \nabla) \bm u(\bm x,t)= -\nabla p(\bm x,t)+\nu\Delta \bm u(\bm x,t)   \eqno{(7)}
$$
$$
\nabla\cdot \bm u(\bm x,t)=0  \eqno{(8)}.
$$   
to the variance of the nonlinear term.   

   In the Fourier transform of Eq. (7)
$$
\frac{\partial u_i(\bm k)}{\partial t} - \mathcal{N}_i(\bm k)= - \nu k^2 u_i(\bm k) \eqno{(9)}
$$
with the nonlinear term
$$
\mathcal{N}_i(\bm k)=-\frac{i}{2}R_{ijm}(\bm k)\iint \delta(\bm k-\bm q-\bm p)u_m(\bm q)u_j(\bm p)d\bm qd\bm p    \eqno{(10)},
$$
$$
R_{ijm}(\bm k)=k_m\delta_{ij}+k_j\delta_{im}-2\frac{k_jk_mk_i}{k^2}  \eqno{(11)}
$$  
the pressure is absent due to the incompressibility Eq. (8).  But this does not mean that the pressure contribution is also absent from the mean-square nonlinearity
$$
\mathcal{N}^2 = \left<|\nabla p(\bm x) +(\bm u(\bm x)\cdot \nabla) \bm u(\bm x)|^2\right>  \eqno{(12)}
$$ 
which spectrum can be defined as
$$
E_\mathcal{N}(k)=4\pi k^2\left<\mathcal{N}_i(-\bm k)\mathcal{N}_i(\bm k))\right>  \eqno{(13)}
$$
and
$$
\mathcal{N}^2 =\int E_\mathcal{N}(k) dk  \eqno{(14)}
$$
It is shown in the Ref \cite{br} that in the frames of the Kolmogorov phenomenology
$$
E_\mathcal{N}(k)\propto \left<|\bm u(\bm x)|^2\right>\varepsilon^{2/3}k^{1/3}  \eqno{(15)}
$$
for the inertial range of scales.
Since in this range scaling of the spectrum of the pressure gradient  \cite{br},\cite{gf} 
$$
 E_{\nabla p}(k)\propto \varepsilon^{4/3}k^{-1/3} \eqno{{(16)}},
$$ 
it was concluded in the Ref. \cite{br} that for large enough Reynolds number the variance of the nonlinearity is dominated by the advection term $(\bm u(\bm x,t)\cdot \nabla) \bm u(\bm x,t)$  (see also Ref. \cite{nt}).

  Let us (dot-) multiply both sides of the Eq. (7) by $\bm u(\bm x,t)$
$$
\frac{1}{2}\frac{\partial \bm u^2}{\partial t}=- \frac{1}{2}(\bm u\cdot \nabla) \bm u^2 -\bm u \cdot \nabla p +\nu~ \bm u\cdot \Delta \bm u   \eqno{(17)}  
$$ 
where $\bm u^2 = \bm u \cdot \bm u$.
 
\begin{figure}
\begin{center}
\includegraphics[width=8cm \vspace{-1.2cm}]{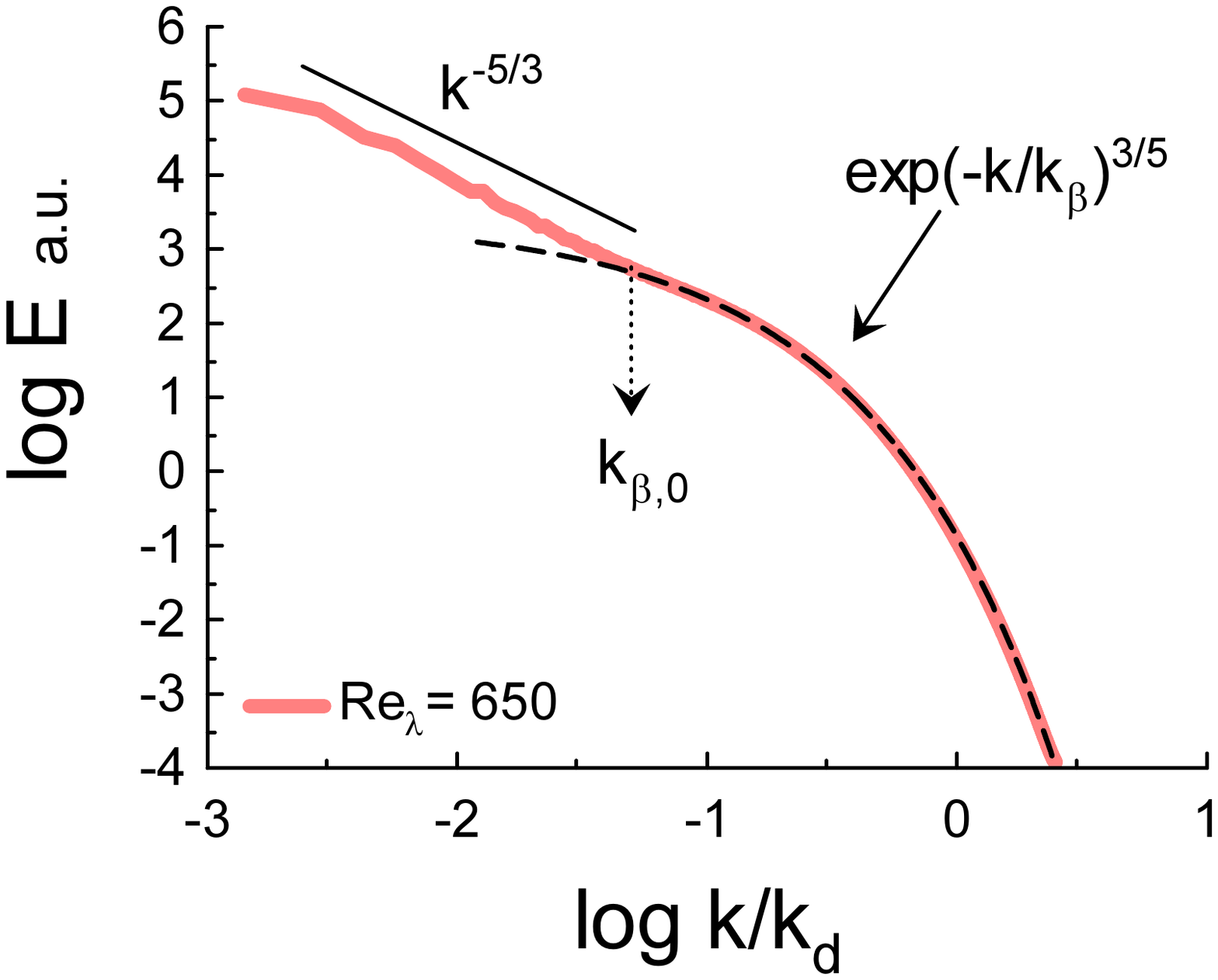}\vspace{-3.5cm}
\caption{\label{fig2} 3D energy spectrum obtained in a recent numerical simulation of the isotropic homogeneous turbulence \cite{isy} at Reynolds number $Re_{\lambda} =650$.   }. 
\end{center}
\end{figure}

 It can be readily shown (in the same manner as it was done for the passive scalar correlation integral - the Corrsin invariant, in Section 15 of the Ref. \cite{my}) that the advection term in the Eq. (17) conserves the energy correlation integral
$$
\mathcal{E}  = \int  \langle {\bf u}^2  \cdot  {\bf u'}^2  \rangle d{\bf r}  \eqno{(18)}
$$ 
This and the domination of the advection term in the inertial range of scales result in adiabatic \cite{ll2} invariance of the energy correlation integral $\mathcal{E}$ in the inertial range of scales. Appearance of the inertial range nearby (at smaller wavenumbers) of the distributed chaos range results in deformation of the energy flux into the distributed chaos range of scales (they were governed by different conservation laws - energy and momentum conservation, respectively). The waves driving the distributed chaos should have a sufficiently fast dynamics in order to adapt themselves to the energy flux from the inertial range. Their scaling dynamics cannot be slower than that in the inertial range in order to do this. Therefore, the energy correlation integral $\mathcal{E}$ should be an adiabatic invariant for these waves as well. This should results in a competition between $\mathcal{E}$ and $I_2$ invariants for governing the scaling Eq. (2). If with developing of the inertial range (with increase of the Reynolds number) the adiabatic invariant $\mathcal{E}$ wins this competition, then
$$
\upsilon (\kappa ) \propto ~\mathcal{E}^{1/4}~\kappa^{3/4} \eqno{(19)}
$$
and from the Eq. (4) one obtains $\beta =3/5$.\\

    Figure 2 shows 3D energy spectrum obtained in a recent numerical simulation of the isotropic homogeneous turbulence \cite{isy} at Reynolds number $Re_{\lambda} =650$ (in the log-log scales, $k_d$ is the Kolmogorov's scale). The solid straight line with slope '-5/3' is drawn to indicate appearance of the Kolmogorov's scaling (inertial) range \cite{my} (cf Fig. 4b of the Ref. \cite{isy}). The dashed line is drawn in order to indicate the distributed chaos spectrum Eq. (3) with the $\beta = 3/5$. The dotted arrow indicates value $k_{\beta , 0}$ separating between the rough (stochastic) and the smooth (chaotic) ranges of wavenumbers: $\ln k_d/ k_{\beta , 0} \simeq 3$ (see the Refs. \cite{b1},\cite{bb}).
    
\section{Buoyancy driven turbulence}

  This consideration can be readily generalized for buoyancy driven turbulence. 
 In the Boussinesq approximation the both stable and unstable stratification can be described by equations
$$ 
\frac{ \partial{\mathbf u}}{\partial t} +{\mathbf u} \cdot \nabla {\mathbf u}  = -\frac{1}{\rho_{0}} \nabla p - N \theta  {\bf e_g} + \nu \nabla^2 {\mathbf u}+{\bf f}  \eqno{(20)}, 
$$ 
$$     
\frac {\partial \theta}{\partial t} +{\mathbf u} \cdot \nabla \theta  = s~N~ {\bf u} \cdot {\bf e_z} + D \nabla^2 \theta \eqno{(21)},
$$  
$$
 \nabla \cdot {\bf u} = 0 \eqno{(22)};
$$
with $s = 1$ for the stable and $s= -1$ for the unstable case and with rescaling the buoyancy field $\theta$ as a velocity. The Brunt-V\"ais\"al\"a frequency $N=\sqrt{-(g/\theta ) (d\bar \theta /dz)}$ for stable stratification (see Ref. \cite{bur} for definition of $N$ for unstable stratification in the Rayleigh-Taylor mixing zone). Usually the unit vector in the buoyancy direction ${\bf e_g} = {\bf e_z}$ (we will consider this standard case). By (dot-) multiplying both sides of the Eq. (20) by ${\mathbf u}$ and both sides of the Eq. (21) by $\theta$, and making their summation one obtains 
$$
\frac{1}{2}\frac{\partial (\bm u^2+ s\theta^2)} {\partial t}=- \frac{1}{2}(\bm u\cdot \nabla) (\bm u^2+ s\theta^2) -\frac{1}{\rho_{0}}\bm u \cdot \nabla p ~ +  
$$  
$$
+ \nu~ \bm u\cdot \Delta \bm u  + sD ~\nabla^2 \theta + {\bf f}  \eqno{(23)}
$$

  At appearance of an inertial range of scales consideration similar to that of the previous section shows that the generalized energy correlation integral
$$
\mathcal{E}_b  = \int \langle ({\bf u}^2 + s ~\theta^2) \cdot  ({\bf u'}^2 + s ~\theta'^2)  \rangle d{\bf r}  \eqno{(24)}
$$ 
is an adiabatic invariant for the waves driving the distributed chaos. And since the dimensionality of the $\mathcal{E}_b $ is the same as dimensionality of the $\mathcal{E}$ one obtains (substituting $\mathcal{E}_b $ in Eq. (19) instead of $\mathcal{E}$) the same value of $\beta =3/5$ in this case as well.\\
\begin{figure}
\begin{center}
\includegraphics[width=8cm \vspace{-1cm}]{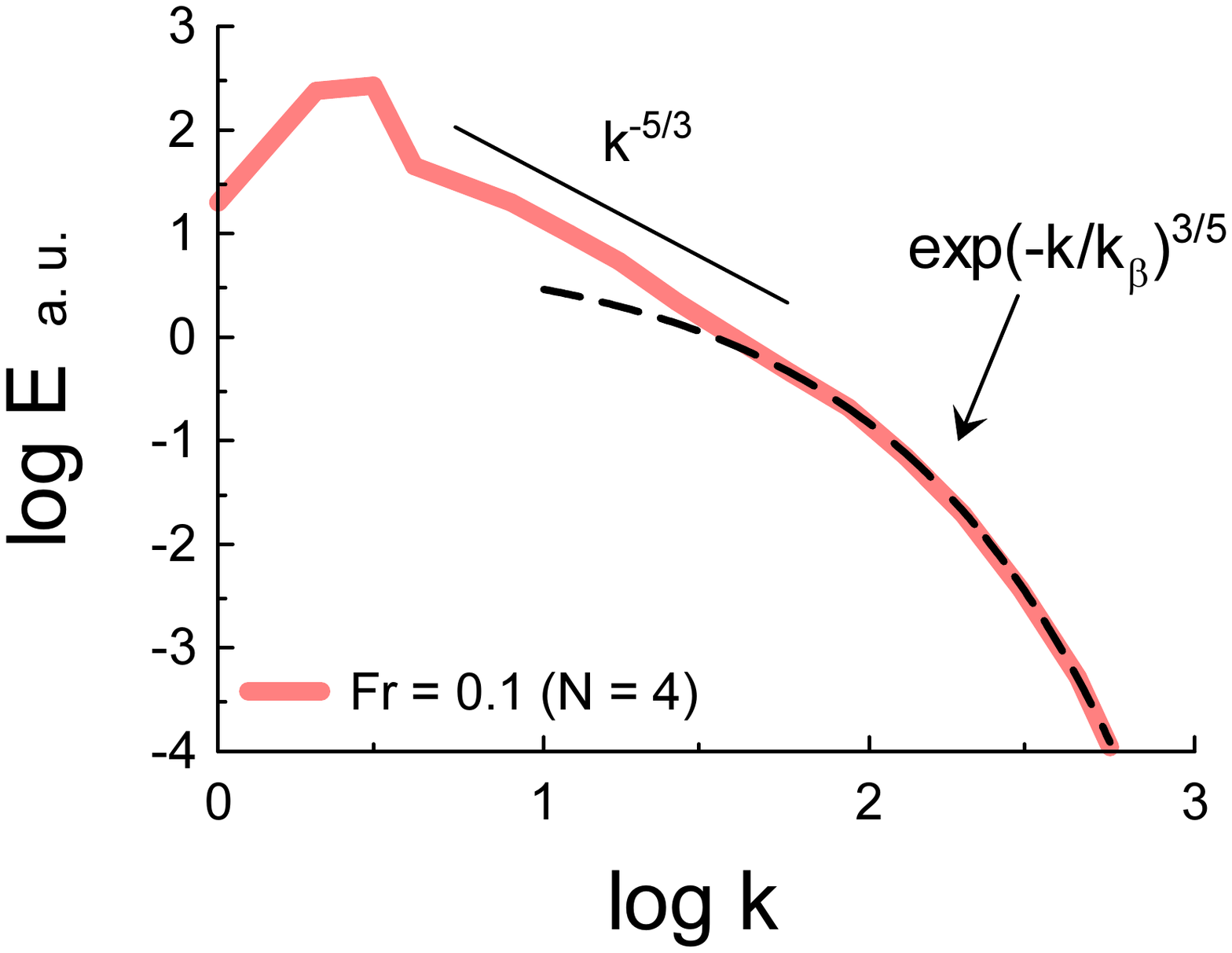}\vspace{-4cm}
\caption{\label{fig3} The generalized energy spectrum ($k$ is isotropic wave number) for $N = 4$. The dashed line is drawn to indicate the spectral law Eq. (3) with $\beta = 3/5$. The straight solid line (with the '-5/3' slope) indicates appearance of the inertial range.  }. 
\end{center}
\end{figure}
\begin{figure}
\begin{center}
\includegraphics[width=8cm \vspace{-1cm}]{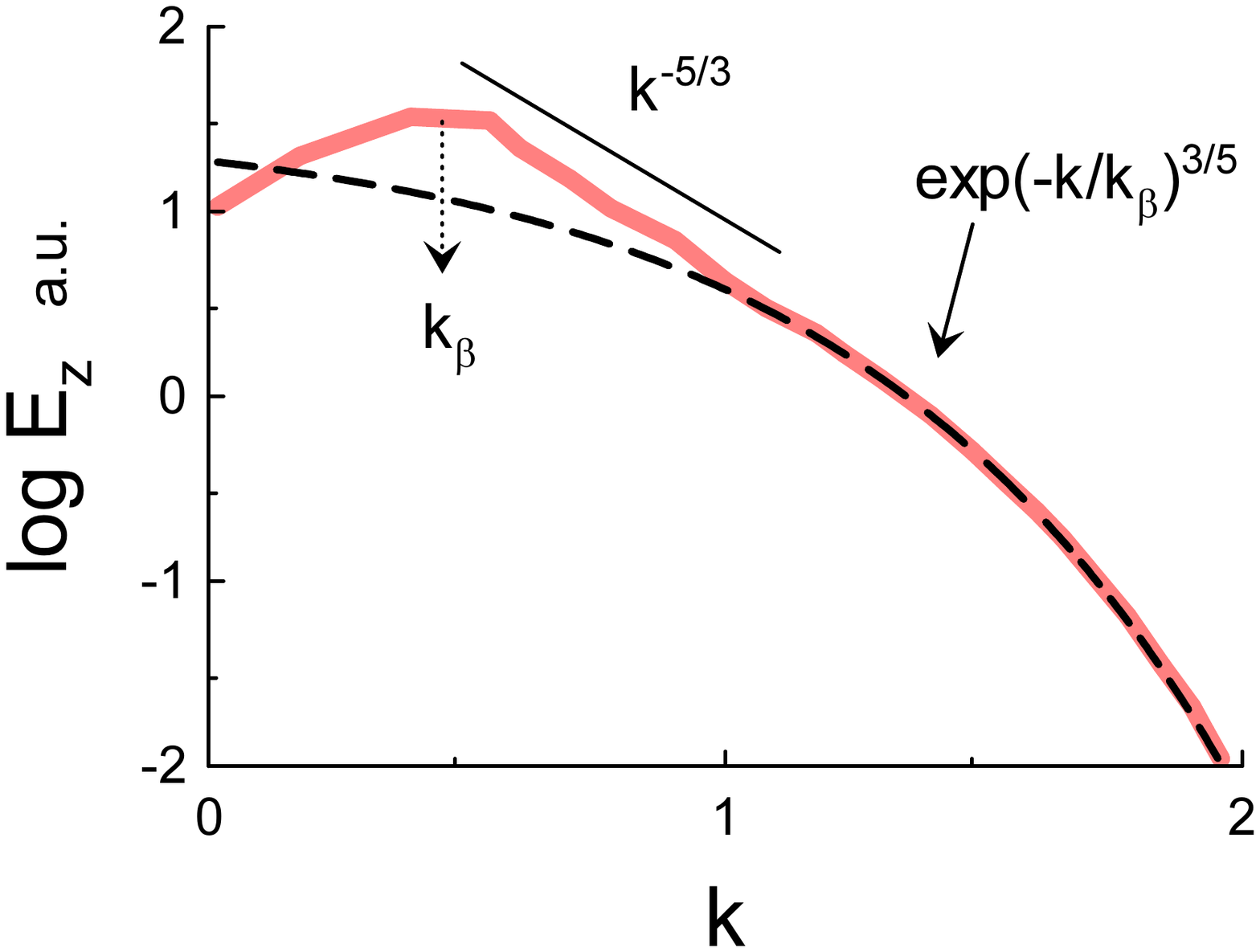}\vspace{-4cm}
\caption{\label{fig4} Power spectrum of the vertical component of velocity field obtained in a DNS for unstable stratification (Rayleigh-Taylor mixing zone) \cite{bof}. The dashed line is drawn to indicate the spectral law Eq. (3) with $\beta = 3/5$. The straight solid line (with the '-5/3' slope) indicates appearance of the inertial range. }. 
\end{center}
\end{figure}
   The spectral data obtained in a DNS for stable stratification with {\it large}-scale forcing were reported in the Ref. \cite{rmp}. This DNS used randomly generated 3D isotropic flows for the initial conditions and for velocity forcing ($Re \simeq 25000$ and $Pr = 1$). The data shown in  Fig. 6a of the Ref. \cite{rmp} were used for presentation in the Fig. 3. Figure 3 shows the generalized energy spectrum ($k$ is isotropic wave number) for $N = 4$ (the Froude number $Fr = 0.1$). The dashed line is drawn to indicate the spectral law Eq. (3) with $\beta = 3/5$. The straight solid line (with the '-5/3' slope) indicates appearance of the inertial range. \\

   Spectral data obtained in a DNS for unstable stratification (Rayleigh-Taylor mixing zone)
were reported in the Ref. \cite{bof}. The data shown in  Fig. 9 of the Ref. \cite{bof} (computed at time $t = 3.1\tau$) were used for presentation of the power spectrum of the vertical component of velocity field in the Fig. 4. The dashed line is drawn to indicate the spectral law Eq. (3) with $\beta = 3/5$. The straight solid line (with the '-5/3' slope) indicates appearance of the inertial range.\\

   The spectra in the inertial range have been corrupted by large-scale structures in the both stable and unstable cases shown in the Figs. 3 and 4, and the presumably inertial ranges are rather short. Nevertheless, the above described mechanism of the inertial range domination in respect of the distributed chaos spectral properties is already clear seen. The large-scale structures, corrupting the inertial range spectrum, do not corrupt the distributed chaos spectrum with its smaller than inertial wavenumbers. The only apparent influence of the large scale structures on the distributed chaos is a tuning of the distributed chaos to these structures in the Rayleigh-Taylor mixing zone, shown by an arrow in the Fig. 4.   

\section{Discussion}

 The energy correlation integral competes (as an adiabatic invariant) not only with the correlation integrals related to the fundamental conservation laws (fundamental symmetries) but also with spontaneous breaking of these symmetries due to the finite-size effects. Therefore the value $\beta =3/5$ competes also with value $\beta = 1/2$ (spontaneous breaking of the space translational symmetry - homogeneity \cite{b2}), for instance. 
 
  \begin{figure}
\begin{center}
\includegraphics[width=8cm \vspace{-1cm}]{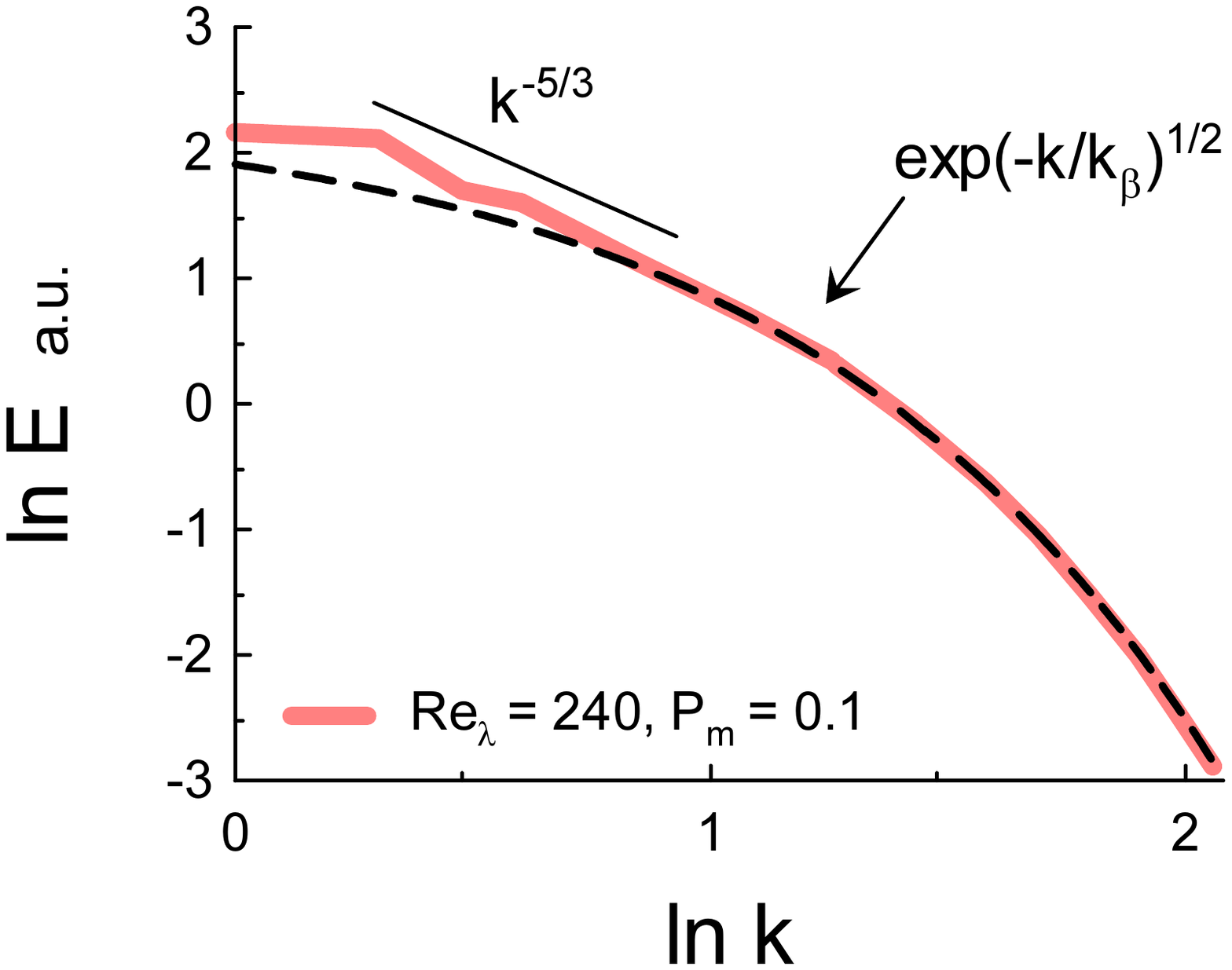}\vspace{-3.5cm}
\caption{\label{fig5} Generalized (total) energy spectrum of a forced (statistically steady) three-dimensional MHD turbulence (the spectral data are from a DNS reported in Fig. 2d of Ref. \cite{g}). }
\end{center}
\end{figure}

    For magnetohydrodynamics (MHD) turbulence the symmetry spontaneous breaking effects are actually dominating ones, because the generalized (total) energy correlation integral is not an adiabatic invariant in this case. Figure 5 shows generalized (total) energy spectrum of a forced (statistically steady) three-dimensional MHD turbulence. The spectral data are from a DNS reported in Fig. 2d of Ref. \cite{g}. The straight solid line (with the '-5/3' slope) indicates appearance of the inertial range, while the dashed line indicates the spectral law Eq. (3) with $\beta =1/2$ corresponding to the spontaneous breaking of the space translational symmetry - homogeneity \cite{b2}).  

\section{Acknowledgement}

I thank D. Fukayama, T. Gotoh, K. P. Iyer, and T. Nakano for sharing their data. I also thank W J. T. Bos and K. P. Iyer for their thought-provoking questions and comments.

\end{document}